\documentclass[fleqn,twoside]{article}
\usepackage{espcrc2}

\usepackage{graphicx}
\usepackage[figuresright]{rotating}

\newcommand{\AmS}{{\protect\the\textfont2
  A\kern-.1667em\lower.5ex\hbox{M}\kern-.125emS}}

\title{Topological Charge Fluctuations and Low-Lying Dirac Eigenmodes}

\author{S.J.~Dong\address[UK]{Department of Physics \& Astronomy, 
        University of Kentucky, Lexington, KY 40506, USA},
        T.~Draper\addressmark,
        I.~Horv\'ath\addressmark\thanks{The speaker at Lattice 2001.},
        N.~Isgur\address[JL]{Jefferson Lab, 12000 Jefferson Avenue,
        Newport News, VA 23606, USA},
        F.X.~Lee\addressmark[JL]\address{Center for Nuclear Studies, 
        George Washington University, Washington, DC 20052, USA},
        J.~McCune\address[UVA]{Department of Physics, University of Virginia,
        Charlottesville, VA 22901, USA},
        J.B.~Zhang\address{CSSM and Dept. of Physics and Math. Physics, 
        University of Adelaide, Adelaide, SA 5005, Australia},
        H.B.~Thacker\addressmark[UVA]}
       
\begin{document}

\begin{abstract}
We discuss the utility of low-lying Dirac eigenmodes for studying the nature
of topological charge fluctuations in QCD. The implications of previous
results using the local chirality histogram method are discussed, and the new
results using the overlap Dirac operator in Wilson gauge backgrounds at 
lattice spacings ranging from $a\approx 0.04\,\mbox{\rm fm}$ 
to $a\approx 0.12\,\mbox{\rm fm}$ are reported. While the degree of local
chirality does not change appreciably closer to the continuum limit, we find
that the size and density of local structures responsible for chiral peaking
do change significantly. The resulting values are in disagreement 
with the assumptions of the Instanton Liquid Model. We conclude that 
the fluctuations of topological charge in the QCD vacuum are not locally 
quantized.
\vspace{1pc}
\end{abstract}

\maketitle
\input epsf

The instanton (semiclassical) picture of the QCD vacuum has undergone 
several refinements since its beginnings as a dilute instanton gas, 
but there are two cornerstones that are inherent in this approach. First, 
it is imagined that the gauge field fluctuates in such a way as to form 
the lumps of approximately quantized topological charge. More precisely, 
it is assumed that there are regions of strong gauge fields that can be 
enclosed by a hypersurface on which the gauge field is approximately 
pure gauge. The topological charge contained in such regions thus have 
approximately quantized values, and we will refer to this situation as 
{\it ``lumpy vacuum''} or vacuum with locally quantized topological charge. 
Secondly, the gauge field in such lumps is assumed to be overwhelmingly 
{\it (anti)self-dual}. Rather successful phenomenology, the Instanton 
Liquid Model (ILM)~\cite{ILM}, has been built around these assumptions.
The basic characteristics of the ILM are the average instanton radius
$\rho\approx 1/3\,\mbox{\rm fm}$ 
and density 
$n\approx 1\,\mbox{\rm fm}^{-4}$ 
with values rather tightly restricted.

Using fermionic response as a signal for gauge fluctuations is a plausible 
strategy on the lattice, since lattice gauge fields are quite rough and 
believed to be affected strongly by artifacts at the scale of the lattice 
spacing. The attemps to eliminate the unphysical part of UV fluctuations 
directly from the gauge field are quite subjective and significantly biased 
as to the local structure of the vacuum they reveal. Indeed, cooling
inevitably leads to lumpy and self-dual gauge fields if global topology is 
preserved in the process. Similarly, smoothing is biased towards the lumpy 
vacuum but not necessarily towards self-duality.

The eigenset of the Dirac operator fully encodes the information about
the topological fluctuations of the underlying gauge field. While relying
on the complete eigenset is impractical, the importance of these fluctuations 
mainly rests in their influence on the propagation of light quarks which,
in turn, depends crucially on the low-lying modes. Hence, it is plausible 
to look for the imprints of topological charge fluctuations in Dirac
near-zero modes~\cite{Hor01A}. This approach actually represents a form 
of {\it smoothing} because the infrared modes are least sensitive to the 
short-distance features atop larger structures in the gauge potential.
However, ``fermionic smoothing'' is determined by the relevant fermionic 
dynamics, thus eliminating the subjective element.

While lattice fermion is not unique (different fermions smooth differently), 
all good actions should give consistent results sufficiently close to the 
continuum limit. However, to make meaningful comparisons, 
the same fermionic action should be used as the lattice spacing changes. 
There appears to be an interesting complementarity here of the type 
``{\it degree of chiral symmetry} $\times$ {\it resolution} $=$ {\it const}''. 
Valid ultralocal actions can not have exact chiral symmetry~\cite{Nultr}, 
but they can have perfect lattice ``resolution'' in the sense that the fermion 
only feels the gauge potential at its location. The actions with exact chiral 
symmetry, however, must be non-ultralocal and their resolution is poor. 
The advantage of the overlap operator over the Wilson-Dirac operator which 
still reflects the global topology quite well and has perfect resolution, 
is thus not obvious {\it a priori}.

A concrete proposal along these lines is the investigation of chiral
orientation parameter $X(n)$ in the low eigenmodes~\cite{Hor01A}. $X(n)$ 
is a local angle in the $|\psi_L|-|\psi_R|$ plane, normalized to lie
in the interval $[-1,1]$. This is quite useful because one can argue that 
the distribution of this quantity over active points on the lattice will 
have a predictable qualitative behaviour when the field is lumpy or self-dual. 
For a lumpy configuration, the emergence of {\it topological near-zero-modes} 
is expected, and the distribution of $X$ should be strongly peaked at the 
extremal values. When the gauge field exhibits extended regions of strong 
self-dual fields, the distribution should tend to be peaked as well. 
We emphasize that these two cases can happen {\it independently}, e.g. the 
lumpy field does not have to be self-dual for the distribution to be peaked. 
Consequently, such a qualitative behaviour is a necessary condition for 
instanton-like vacuum but by no means sufficient! 

The first study of local chirality with Wilson fermions on Wilson gauge
backgrounds at lattice spacing $a\!\approx\!0.17\,\mbox{\rm fm}$ found 
results inconsistent with the instanton-like vacuum~\cite{Hor01A}. 
Later works with Wilson fermions at smaller lattice spacing and also with 
the overlap revealed however, that a visible degree of double-peaked 
behaviour is present~\cite{Followup}. Our overlap calculation on Wilson 
gauge fields at $a\!\approx\!0.123,\,0.042\,\mbox{\rm fm}$ gave the following
results

\smallskip
\epsfxsize .85\hsize
\centerline{\epsffile {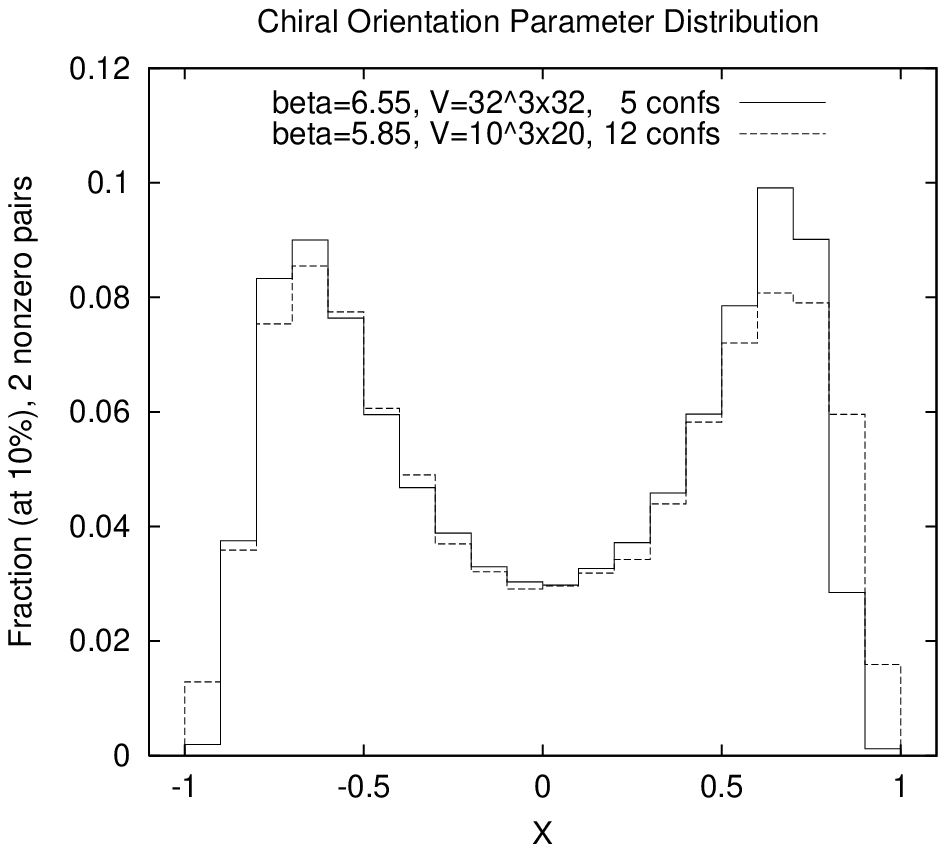}}
\smallskip

\noindent indicating that there does not appear to be a significant change 
in the degree of local chirality closer to the continuum limit. While
the observed peaking does not imply the dominance of instantons, it does 
force us to look for quantitative comparisons rather than qualitative ones. 
It might be indeed quite revealing to compare the probability distribution 
of $X$ for an instanton liquid ensemble and for QCD using the overlap 
operator. To illustrate this point, we compare our results to the prediction 
of the dilute instanton gas model,

\smallskip
\epsfxsize .85\hsize
\centerline{\epsffile {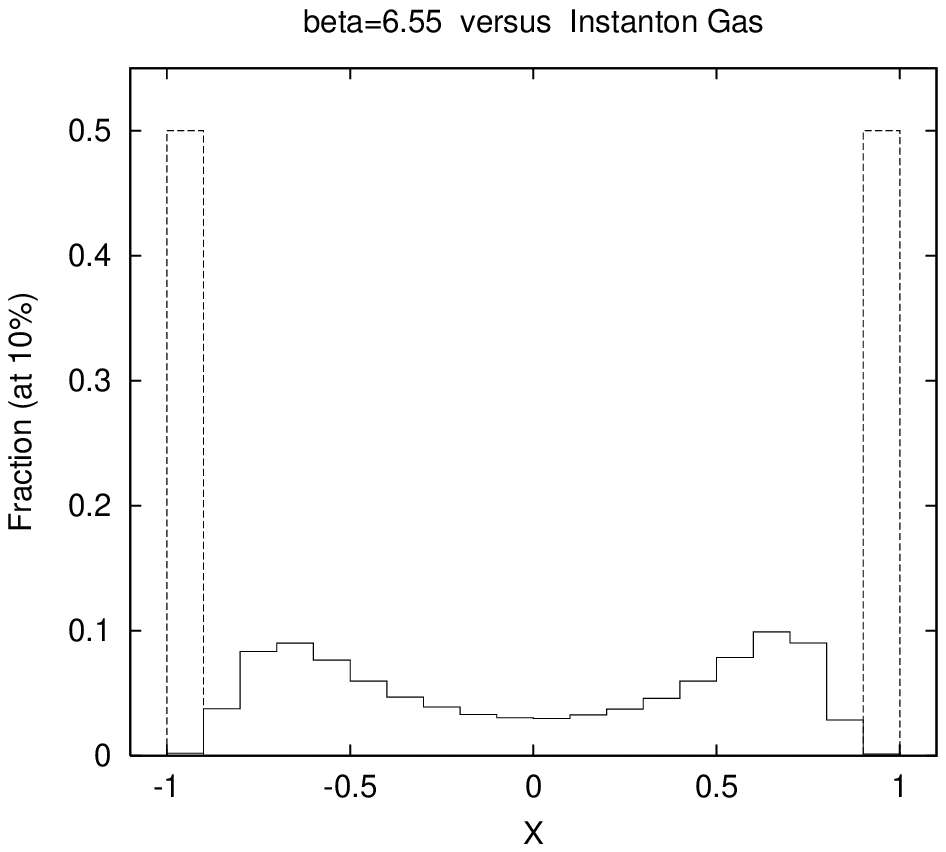}}
\smallskip

\noindent confirming that it is not applicable.

If the low-lying mode arises from mixing of t'Hooft ``would be'' zero modes 
associated with instantons, then the regions of coherent local chirality should
be of a particular size and abundance. To check whether the behaviour of
$X$-distribution can be ascribed to gauge structures with ILM parameters, we 
have found the local maxima over the distance $\sqrt3$ of density ($\psi^+\psi$),
and selected ones that (1) decay on average along the eight axis directions 
through distance $\sqrt3$ (this excludes accidental maxima and those not 
likely due to instantons), and (2) contribute to the peaks of $X$-distribution. 
We then determined the size of selected structures as (a) the radius of the largest 
hypersphere centered at the maximum and containing the sites with the same sign 
of chirality ($\psi^+\gamma_5\psi$), and (b) the radius of the largest hypersphere 
within which density doesn't fall below $1/8$ of the center value (this 
corresponds to the radius in the case of an instanton). Using the first nonzero 
modes we found

\smallskip
\epsfxsize .85\hsize
\centerline{\epsffile {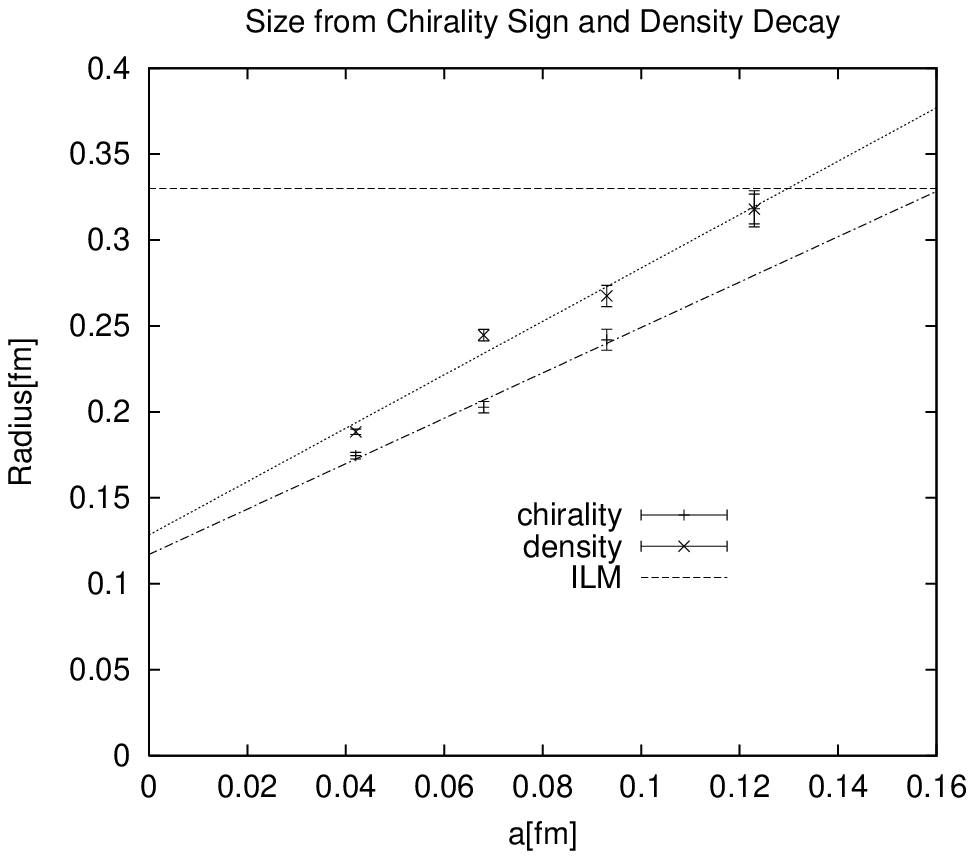}}
\smallskip

\noindent with values significantly below the ILM assumption close to continuum
limit. Similarly, for the density of these structures we have obtained 

\smallskip
\epsfxsize .85\hsize
\centerline{\epsffile {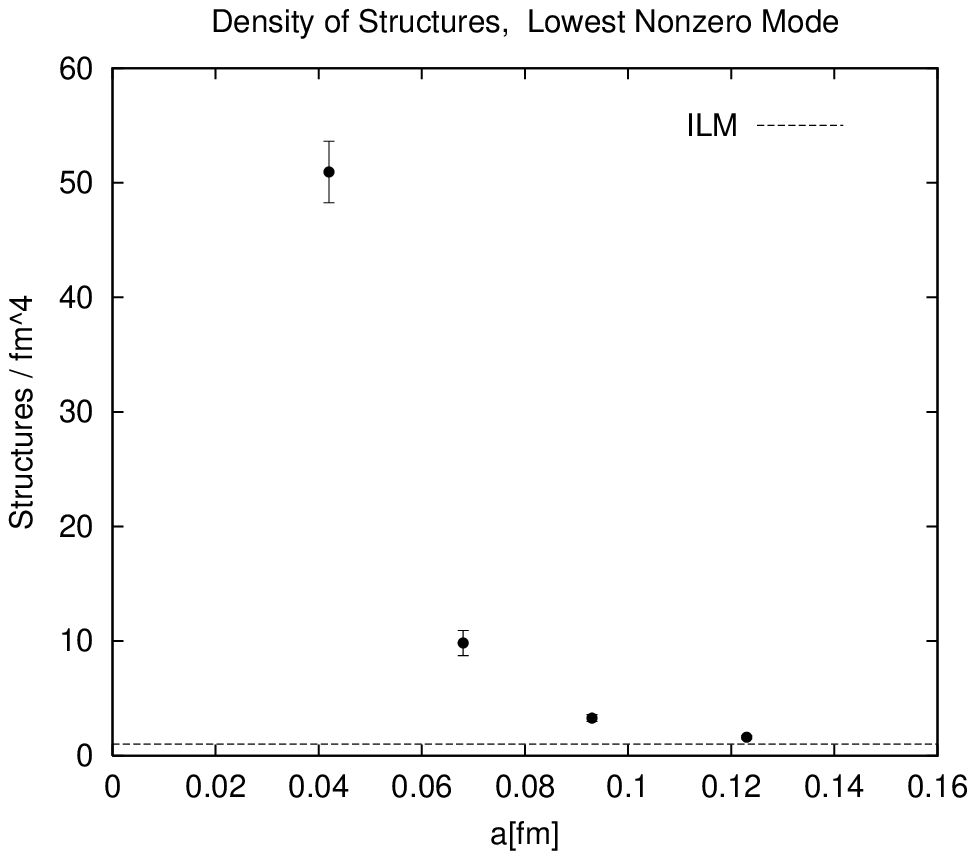}}
\smallskip

\noindent in a marked disagreement with the ILM. A few points need to be emphasized 
here. (A) There should not be any additional smoothing or other manipulation done 
on the low-lying eigenmodes. The lowest nonzero infrared mode provides a 
maximally smoothed imprint of topological charge fluctuations as seen by the 
propagating fermion. (B) The typical ``size'' of the chirality correlator in the 
lowest modes (representing an upper bound on the average size of underlying 
individual structures) at $a\!\approx\!0.042\,\mbox{\rm fm}$ is about 
$0.23\, \mbox{\rm fm}$. 
(C) The possibility that the large density of structures we find is due to 
dislocations is unlikely and can be excluded in several ways.

How do we interpret our data? They are easy to understand if one abandons the 
assumption that topological charge in QCD vacuum is approximately locally
quantized. For quantized topological charges, the calculable occurence of 
$0.15\,\mbox{\rm fm}$ instantons should be negligible, but there is no such 
restriction for inhomogeneous fluctuations without quantization. Similarly, if 
the large number of structures we found corresponded to quantized ``lumps'', then
the typical values of global topological charge (and susceptibility) would be much
higher than we observe, but not necessarily so without quantization. We conclude 
that it is unlikely that the observed peaking of local chirality distribution is
due to instanton-like objects with the parameters of the ILM. Instead, we find
an evidence for serious breakdown of the semiclassical picture of the 
QCD vacuum~\cite{Witten}.

\end{document}